\documentclass[twocolumn, showpacs, nofootinbib, aps, prd]{revtex4}

\usepackage{graphicx}
\usepackage{amsfonts}

\begin{document}

\title{Revisiting the Pion Leading-Twist Distribution Amplitude within the QCD Background Field Theory}

\author{Tao Zhong$^{1}$}
\email{zhongtao@ihep.ac.cn}
\author{Xing-Gang Wu$^{2}$}
\email{wuxg@cqu.edu.cn}
\author{Zhi-Gang Wang$^{3}$}
\email{zgwang@aliyun.com}
\author{Tao Huang$^1$}
\email{huangtao@ihep.ac.cn}
\author{Hai-Bing Fu$^{2}$}
\author{Hua-Yong Han$^{2}$}

\address{$^1$ Institute of High Energy Physics and Theoretical Physics Center for Science Facilities, Chinese Academy of Sciences, Beijing 100049, P.R. China \\
$^{2}$ Department of Physics, Chongqing University, Chongqing 401331, P.R. China\\
$^{3}$ Department of Physics, North China Electric Power University, Baoding 071003, P. R. China}

\date{\today}

\begin{abstract}

We study the pion leading-twist distribution amplitude (DA) within the framework of SVZ sum rules under the background field theory. To improve the accuracy of the sum rules, we expand both the quark propagator and the vertex $(z\cdot \tensor{D})^n$ of the correlator up to dimension-six operators in the background field theory. The sum rules for the pion DA moments are obtained, in which all condensates up to dimension-six have been taken into consideration. Using the sum rules, we obtain $\left<\xi^2\right>|_{\rm 1\;GeV} = 0.338 \pm 0.032$, $\left<\xi^4\right>|_{\rm 1\;GeV} = 0.211 \pm 0.030$ and $\left<\xi^6\right>|_{\rm 1\;GeV} = 0.163 \pm 0.030$. It is shown that the dimension-six condensates shall provide sizable contributions to the pion DA moments. We show that the first Gegenbauer moment of the pion leading-twist DA is $a^\pi_2|_{\rm 1\;GeV} = 0.403 \pm 0.093$, which is consistent with those obtained in the literature within errors but prefers a larger central value as indicated by lattice QCD predictions.

\end{abstract}


\maketitle

\section{introduction}

The pion distribution amplitude (DA) is an important element for applying the QCD factorization theory and QCD light-cone sum rules (LCSR) to exclusive processes involving pion. For examples, it is important for understanding the semi-leptonic decays $B\to \pi l \nu$ and $D\to \pi l \nu$, the pion-photon transition form factor (TFF) $F_{\pi\gamma}(Q^2)$, the exclusive process $B\to \pi\pi$, and etc.. Inversely, one can make use of all those processes to obtain stringent constraints on the pion DA~\cite{INVDA1,INVDA2}. Because of its simplest structure, the pion leading-twist DA has attracted much attention after the pioneering works done by Refs.\cite{ASDA,LDA,CZDA}. Unfortunately, at present, there is no definite conclusion on the behavior of pion leading-twist DA. For example, the BABAR measurement for the pion-photon TFF~\cite{BABAR} supports the Chernyak-Zhitnitsky (CZ)-like~\cite{CZDA} behavior; while the corresponding data by the Belle Collaboration~\cite{BELLE} supports the asymptotic-like behavior~\cite{ASDA}.

The pion leading-twist DA at the scale $\mu$ can be expanded into a Gegenbauer polynomial as~\cite{GENDA},
\begin{eqnarray}
\phi_\pi (\mu, x) = 6x(1-x) \left[ 1 + \sum^\infty_{n=2} a^\pi_n(\mu) C^{3/2}_n(2x-1) \right], \label{DAgegexp}
\end{eqnarray}
where the $C^{3/2}_n(2x-1)$ are Gegenbauer polynomials and the $a^\pi_n(\mu)$ are Gegenbauer moments. Theoretically, the Gegenbauer moments have been calculated by QCD sum rules~\cite{PDA_SR1,PDA_SR2,PDA_SR3,PDA_SR4,PDA_SR5} and the lattice gauge theory~\cite{PDA_L1,PDA_L2,PDA_L3,PDA_L4}. Phenomenologically, the Gegenbauer moments have also been estimated from the pion-photon transition form factor~\cite{PDA_LC1,PDA_LC21,PDA_LC22,PDA_LC23}, the pion electromagnetic form factor~\cite{PDA_LC3,PDA_LC4}, the $B\to\pi$ TFFs~\cite{PDA_LC5,PDA_LC6}, and etc. Most of those estimations are consistent with each other within errors. It is noted that the central values of them are different from each other, especially, the lattice QCD estimations are bigger than those of QCD sum rules. Due to the forthcoming more accurate data, it is helpful to provide a more accurate theoretical prediction on the pion DA for a better comparison with the data.

Since its invention~\cite{SVZ}, the Shifman-Vainshtein-Zakharov (SVZ) sum rules has been widely adopted for dealing with the hadron phenomenology. In addition, the background field theory provides a systematic description for its key components, i.e. the vacuum condensates, from the viewpoint of QCD field theory~\cite{BG1,BG2,BG3,BG4,BG_HT,pro_func}. It assumes that the quark and gluon fields are composed of the background fields and the quantum fluctuations around them. Thus, to take the background field theory as the foundation for the QCD sum rules, it not only has distinct physical picture, but also greatly simplifies the calculation due to its capability of adopting different gauges for quantum fluctuations and background fields. The QCD sum rules within the background field theory has already been adopted to study the meson properties, cf.Refs.\cite{XIANG,XIANG2,XINGHUA,ZHONG1,ZHONG2,HUAYONG}.

Previous sum rule estimations for the pion DA moments are usually done up to dimension-four condensates, or including part of the dimension-six condensates' contributions. In the present paper, we shall adopt the SVZ sum rules under the background field theory to make a detailed study on the pion leading-twist DA up to dimension-six operators. For the purpose, as the first time, we shall expand both the quark propagator and the vertex operator $(z\cdot \tensor{D})^n$ up to a more complex form with full dimension-six operators' contributions so as to achieve an accurate QCD sum rule estimation. We shall show that those dimension-six operators shall result in new dimension-six condensates that do provide sizable contributions and should be taken into consideration for a sound estimation, i.e. they shall bring sizable changes in our previous predictions on the pion DA behavior.

The remaining parts of the paper are organized as follows. In section II, we first present a brief introduction of the QCD sum rules and then provide the quark propagator and the vertex $(z\cdot \tensor{D})^n$ up to dimension-six operators under the background field theory. In section III, we present the calculation technology for deriving the sum rules for pion leading-twist DA moments. Numerical results are given in section IV. Section V is reserved for a summary. For convenience, formulas for the propagators and the vacuum matrix elements in $D$-dimension space are presented in the appendix.

\section{quark propagator, vertex operator under the background field theory}

Within the SVZ sum rules, the hadrons are represented by their interpolating quark currents taken at large virtualities. The correlator of those currents can be treated within the operator product expansion, whose short-distance coefficients are calculated using QCD perturbation theory, whereas its long-distance parts are written according to the non-perturbative but universal vacuum condensates, such as the quark condensate $\left<\bar{q}q\right>$, the gluon condensate $\left<G^2\right>$, and etc. Furthermore, under the background field theory, the quark and gluon fields are composed of background fields and quantum fluctuations around them. Because of the influence from background fields, the quark and gluon propagators shall include nonperturbative component inevitably, which however can be treated via a systematic way.

Within the background field theory, the gluon field $\mathcal{A}^A_\mu(x)$ and quark field $\psi(x)$ in the QCD Lagrangian or Green function should be replaced by
\begin{eqnarray}
\mathcal{A}^A_\mu(x) &\to& \mathcal{A}^A_\mu(x) + \phi^A_\mu(x), \label{bfrep0} \\
\psi(x) &\to& \psi(x) + \eta(x), \label{bfrep}
\end{eqnarray}
where in the right hand side of the arrow, $\mathcal{A}^A_\mu(x)$ with $(A = 1, \cdots, 8)$ and $\psi(x)$ indicate the background fields of the gluon and quark, respectively. $\phi^A_\mu(x)$ and $\eta(x)$ stand for the gluon and quark's quantum field, i.e., the quantum fluctuation on the background fields. The Lagrangian with the background field theory can be found in Ref.\cite{BG_HT}. Usually, the background quark and gluon fields satisfy the following equations of motion
\begin{eqnarray} &&
(i \slash \!\!\!\! D - m)\psi(x) = 0, \nonumber \\ &&
\widetilde{D}^{AB}_\mu G^{B\nu\mu}(x) = g_s \bar{\psi}(x) \gamma^\nu T^A \psi(x),
\label{motequ}
\end{eqnarray}
where $D_\mu = \partial_\mu - ig_s T^A \mathcal{A}^A_\mu(x)$ and $\widetilde{D}^{AB}_\mu = \delta^{AB} - g_s f^{ABC} \mathcal{A}^C_\mu(x)$ indicate the fundamental and adjoint representations of the gauge covariant derivatives respectively. As an advantage of the background field theory, one can take different gauges for dealing with the quantum fluctuations and background fields. More specifically, one can adopt the ``background gauge"~\cite{BG1,BG2,BG3,BG4}
\begin{eqnarray}
\widetilde{D}^{AB}_\mu \phi^{B \mu}(x) = 0, \label{bfgau}
\end{eqnarray}
for the gluon quantum field, the Schwinger gauge or the ``fixed-point gauge"~\cite{FPgau}
\begin{eqnarray}
x^\mu \mathcal{A}^A_\mu(x) = 0, \label{fpgau}
\end{eqnarray}
for the background field.

\subsection{Quark propagator under background field theory}

\begin{figure*}
\includegraphics[width=0.85\textwidth]{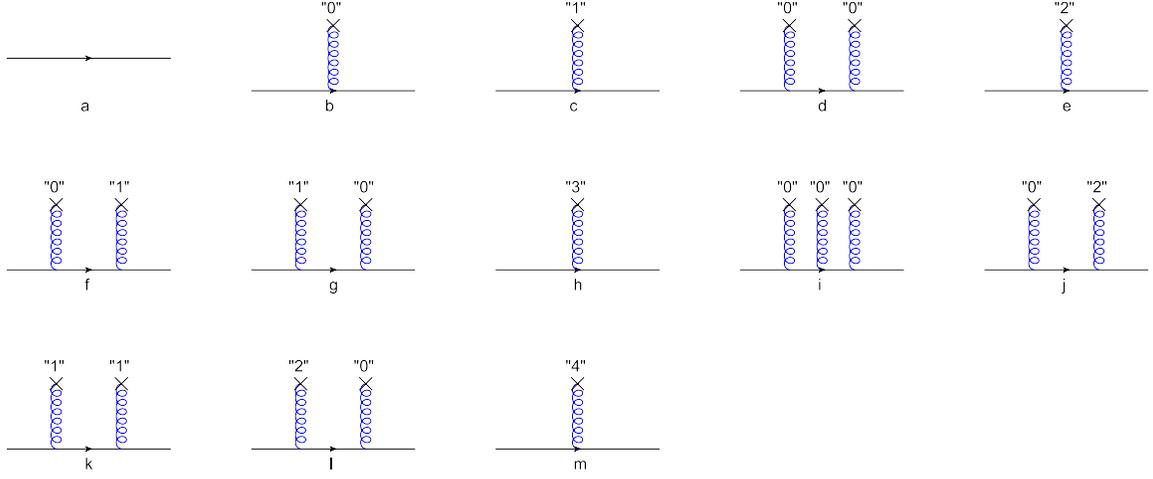}
\caption{Feynman diagrams for the quark propagator within the background field theory which shall results in operators up to dimension-six. The cross ($\times$) attached to the gluon line indicates the tensor of the local gluon background field, in which ``$n$" stands for $n$-th order covariant derivative.} \label{propagator}
\end{figure*}

Within the framework of the background field theory, the quark propagator would be affected by the background quark and/or gluon fields, which satisfies the equation
\begin{eqnarray}
(i \slash\!\!\!\! D - m) S_F(x,0) = \delta^4 (x).
\label{quaprofun1}
\end{eqnarray}
If taking
\begin{eqnarray}
S_F(x,0) = (i\slash\!\!\!\! D + m) \mathcal{D}(x,0),
\label{quaprofun2}
\end{eqnarray}
Eq.(\ref{quaprofun1}) can be changed as
\begin{eqnarray}
(\Box - \mathcal{P}_\mu \partial^\mu - \mathcal{Q} + m^2) \mathcal{D}(x,0) = \delta^4(x),
\label{quaprofun3}
\end{eqnarray}
where $\Box = \partial^2$, and
\begin{eqnarray}
\mathcal{P}_\mu &=& 2i\mathcal{A}_\mu(x), \nonumber\\
\mathcal{Q} &=& \gamma^\nu \gamma^\mu \mathcal{A}_\nu(x) \mathcal{A}_\mu(x) + i \gamma^\nu \gamma^\mu \partial_\nu \mathcal{A}_\mu(x). \nonumber
\end{eqnarray}

Moreover, using the ``fixed-point gauge" as shown in Eq.(\ref{fpgau}), the gluon background field can be expressed by the gauge invariant $G_{\mu\nu;\alpha_1 \cdots \alpha_n}$ as
\begin{widetext}
\begin{eqnarray}
\mathcal{A}_\mu(x) &=& \frac{1}{2} x^\nu G_{\nu\mu} + \frac{1}{3} x^\nu x^\alpha G_{\nu\mu;\alpha} + \frac{1}{8} x^\nu x^\alpha x^\beta G_{\nu\mu;\alpha\beta}
+ \frac{1}{30} x^\nu x^\alpha x^\beta x^\gamma G_{\nu\mu;\alpha\beta\gamma}
+ \frac{1}{144} x^\nu x^\alpha x^\beta x^\gamma x^\delta G_{\nu\mu;\alpha\beta\gamma\delta} + \cdots, \label{gluexp}
\end{eqnarray}
\end{widetext}
where $G_{\mu\nu;\alpha_1 \cdots \alpha_n}$ is a notation for $(D_{\alpha_1}\cdots D_{\alpha_n}G_{\mu\nu})(0)$, where the indexes $\alpha_1 \cdots \alpha_n$ indicate the covariant derivative up to $n$-th order. $G_{\mu\nu}=G^a_{\mu\nu}T^a$ with $G^a_{\mu\nu}= \partial_\mu A^a_\nu - \partial_\nu A^{a}_\mu + g f^{abc} A^b_\mu A^c_\nu$ stands for the gluon field strength tensor and $f^{abc} (a,b,c = 1,2,\cdots ,8)$ is the structure constant of $SU(3)$ group.

Substituting Eq.(\ref{gluexp}) into Eq.(\ref{quaprofun3}), we obtain the expressions for $\mathcal{D}(x,0)$. By further using Eq.(\ref{quaprofun2}), we can obtain the required quark propagator in the background field, which can be expressed as
\begin{widetext}
\begin{equation}
S_F(x,0)=S_F^0(x,0)+S_F^2(x,0)+S_F^3(x,0)+ \sum_{i=1}^{2} S_F^{4(i)}(x,0) + \sum_{i=1}^{3} S_F^{5(i)}(x,0) + \sum_{i=1}^{5} S_F^{6(i)}(x,0). \label{prop4}
\end{equation}
\end{widetext}
The quark propagators with various gauge invariant tensors $G_{\mu\nu;\alpha_1 \cdots \alpha_n}$ that shall result in up to dimension-six operators are presented in Appendix A. The Feynman diagrams for the quark propagators (\ref{appA1},$\cdots$,\ref{prop2}) that with various gauge invariant tensors are shown in Fig.(\ref{propagator}), where thirteen figures, i.e. Fig.(\ref{propagator}a), $\cdots$, Fig.(\ref{propagator}m), correspond to $S_F^{0}(x,0)$, $\cdots$, $S_F^{6(5)}(x,0)$, respectively. The cross ($\times$) attached to the gluon line indicates the tensor of the local gluon background field with ``$n$" stands for $n$-th order covariant derivative.

Because the ``fixed-point gauge", as indicated by Eq.(\ref{fpgau}), violates the translation invariance, the quark propagator from $x$ to $0$, $S_F(0,x)$, can not be directly obtained by applying the replacement $x \to -x$ in Eq.(\ref{prop4}). It can be related with $S_F(x,0)$ via the relation~\cite{REVPRO}
\begin{eqnarray}
S_F(0,x|\mathcal{A}) = C S^T_F(x,0|-\mathcal{A}^T) C^{-1},
\label{revpro}
\end{eqnarray}
where $C$ stands for the charge conjugation matrix and the symbol $T$ indicates transpose of both the Dirac and the color matrices.

\subsection{Vertex operator under background field theory}

\begin{figure*}
\includegraphics[width=0.85\textwidth]{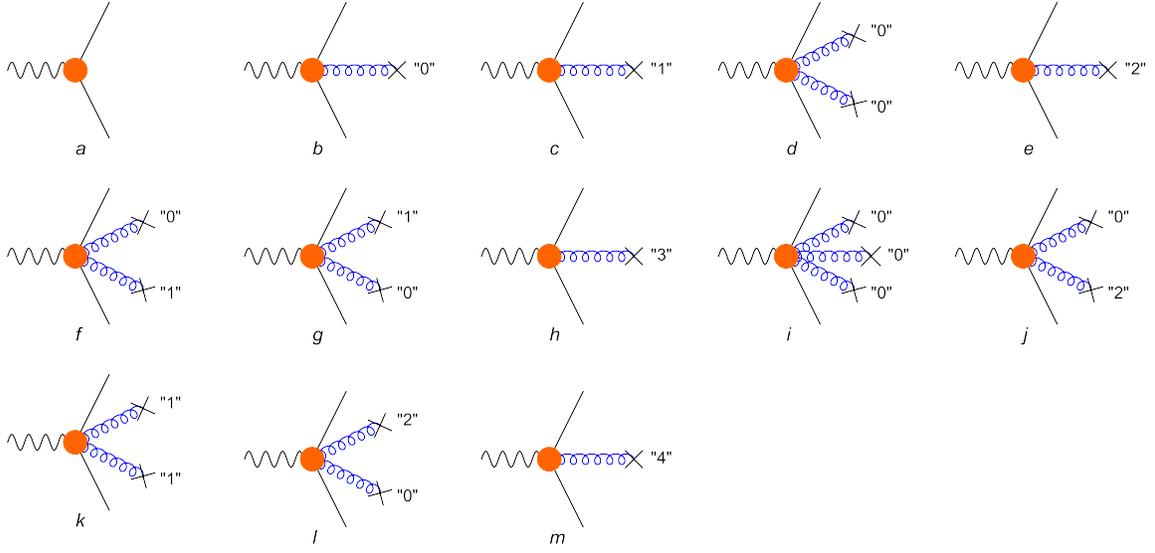}
\caption{Feynman diagrams for the vertex operator $\Gamma (z\cdot \tensor{D})^n$ under the background field theory up to dimension-six. The cross ($\times$) attached to the gluon line indicates the tensor of the local gluon background field, in which ``$n$" stands for $n$-th order covariant derivative.}
\label{vertex}
\end{figure*}

When applying the SVZ sum rules to calculate the moments of meson, one would encounter the vertex operator, $\Gamma(z\cdot \tensor{D})^n$, where $\Gamma$ indicates some kind of Dirac matrices, e.g., $\Gamma = \not\! z \gamma_5$ for the pion leading-twist DA, $\Gamma = \gamma_5$ or $\sigma_{\mu\nu} \gamma_5$ for the pion twist-3 DAs, and etc.. We have
\begin{eqnarray}
(z\cdot \tensor{D})^n = (z\cdot \overrightarrow{D} - z\cdot \overleftarrow{D})^n = (z\cdot \tensor{\partial} + z\cdot B)^n + \cdots,
\end{eqnarray}
where the $\cdots$ stands for the higher-order terms which are irrelevant for our present analysis and
\begin{eqnarray}
z\cdot B &=& -2 i z\cdot \mathcal{A} \nonumber\\
         &=& -i x^\mu z^\nu G_{\mu\nu} - \frac{2i}{3} x^\mu x^\rho z^\nu G_{\mu\nu;\rho} \nonumber\\
&& - \frac{i}{4} x^\mu x^\rho x^\sigma z^\nu G_{\mu\nu;\rho\sigma} - \frac{i}{15} x^\mu x^\rho x^\sigma x^\lambda z^\nu G_{\mu\nu;\rho\sigma\lambda} \nonumber\\
&& - \frac{i}{72} x^\mu x^\rho x^\sigma x^\lambda x^\tau z^\nu G_{\mu\nu;\rho\sigma\lambda\tau} + \cdots. \label{zB}
\end{eqnarray}

We can expand the operator $(z\cdot \tensor{D})^n$ into series with operators $(z\cdot \tensor{\partial})^n$ and $G_{\mu\nu;\rho\cdots}$. For the purpose, we first expand $(z\cdot \tensor{D})^n$ as
\begin{eqnarray}
(z\cdot \tensor{D})^0 &=& 1, \nonumber\\
(z\cdot \tensor{D})^1 &=& z\cdot \tensor{\partial} + z\cdot B, \nonumber\\
(z\cdot \tensor{D})^2 &=& (z\cdot \tensor{\partial})^2 + 2(z\cdot \tensor{\partial}) (z\cdot \underline{B}) + (z\cdot B)^2, \nonumber\\
(z\cdot \tensor{D})^3 &=& (z\cdot \tensor{\partial})^3 + 3(z\cdot \tensor{\partial})^2 (z\cdot \underline{B}) + \left[ (z\cdot \partial)^2 (z\cdot B) \right] \nonumber\\
&& + 3 (z\cdot \tensor{\partial}) (z\cdot \underline{B})^2 + (z\cdot B)^3, \nonumber\\
&& \cdots \cdots , \nonumber
\end{eqnarray}
where the ``underline" below $B$ (or the latter $x$) indicates that the operation $\tensor{\partial}$ does not act on it. In deriving those equations, the following equation has been adopted,
\begin{displaymath}
(z\cdot \tensor{\partial})^n (z\cdot B) = \sum^n_{k=0} \frac{n!}{k! (n-k!)} (z\cdot \tensor{\partial})^{n-k} \left[ (z\cdot \partial)^k (z\cdot \underline{B}) \right].
\end{displaymath}
By keeping only those terms that shall leads to operators up to dimension-six, we obtain
\begin{eqnarray}
(z\cdot \tensor{D})^n_0 &=& (z\cdot \tensor{\partial})^n, \nonumber\\
(z\cdot \tensor{D})^n_2 &=& -i \times n (z\cdot \tensor{\partial})^{n-1} \underline{x}^\mu z^\nu G_{\mu\nu}, \nonumber\\
(z\cdot \tensor{D})^n_3 &=& -\frac{2i}{3} \times n (z\cdot \tensor{\partial})^{n-1} \underline{x}^\mu \underline{x}^\rho z^\nu G_{\mu\nu;\rho}, \nonumber\\
(z\cdot \tensor{D})^n_{4(1)} &=& -\frac{n(n-1)}{2} (z\cdot \tensor{\partial})^{n-2}
\underline{x}^\mu \underline{x}^\rho z^\nu z^\sigma G_{\mu\nu} G_{\rho\sigma},
\nonumber\\
(z\cdot \tensor{D})^n_{4(2)} &=& \left[ -\frac{i}{4} \times n (z\cdot \tensor{\partial})^{n-1} \underline{x}^\mu \underline{x}^\rho \underline{x}^\sigma z^\nu
\right. \nonumber\\
&& - \frac{i}{12} n(n-1)(n-2) (z\cdot \tensor{\partial})^{n-3}
\nonumber\\
&& \left. \times \underline{x}^\mu z^\rho z^\sigma z^\nu \right] G_{\mu\nu;\rho\sigma},
\nonumber\\
(z\cdot \tensor{D})^n_{5(1)} &=& -\frac{n(n-1)}{3} (z\cdot \tensor{\partial})^{n-2}
\nonumber\\
&&\times \underline{x}^\mu \underline{x}^\rho \underline{x}^\lambda z^\nu z^\sigma G_{\mu\nu} G_{\rho\sigma;\lambda}, \nonumber\\
(z\cdot \tensor{D})^n_{5(2)} &=& -\frac{n(n-1)}{3} (z\cdot \tensor{\partial})^{n-2}
\nonumber\\
&& \times \underline{x}^\mu \underline{x}^\rho \underline{x}^\lambda z^\nu z^\sigma G_{\mu\nu;\lambda} G_{\rho\sigma}, \nonumber\\
(z\cdot \tensor{D})^n_{5(3)} &=& \left[ -\frac{i}{15} \times n (z\cdot \tensor{\partial})^{n-1} \underline{x}^\mu \underline{x}^\rho \underline{x}^\sigma \underline{x}^\lambda z^\nu \right. \nonumber\\
&& - \frac{i}{45} n(n-1)(n-2) (z\cdot \tensor{\partial})^{n-3} \nonumber\\
&& \left. \times \underline{x}^\mu \left( \underline{x}^\rho z^\sigma z^\lambda + z^\rho \underline{x}^\sigma z^\lambda + z^\rho z^\sigma \underline{x}^\lambda \right) z^\nu \right] \nonumber\\
&& \times G_{\mu\nu;\rho\sigma\lambda}, \nonumber\\
(z\cdot \tensor{D})^n_{6(1)} &=& \frac{i}{6} n(n-1)(n-2) (z\cdot \tensor{\partial})^{n-3}
\nonumber\\
&& \times \underline{x}^\mu \underline{x}^\rho \underline{x}^\lambda z^\nu z^\sigma z^\tau G_{\mu\nu} G_{\rho\sigma} G_{\lambda\tau}, \nonumber\\
(z\cdot \tensor{D})^n_{6(2)} &=& \left[ -\frac{n(n-1)}{8} (z\cdot \tensor{\partial})^{n-2} \underline{x}^\mu \underline{x}^\rho \underline{x}^\lambda \underline{x}^\tau z^\nu z^\sigma \right. \nonumber\\
&& - \frac{n(n-1)(n-2)(n-3)}{12} (z\cdot \tensor{\partial})^{n-4} \nonumber\\
&& \left. \times \underline{x}^\mu \underline{x}^\rho z^\lambda z^\tau z^\nu z^\sigma \right] G_{\mu\nu} G_{\rho\sigma;\lambda\tau}, \nonumber\\
(z\cdot \tensor{D})^n_{6(3)} &=& -\frac{2}{9} n(n-1) (z\cdot \tensor{\partial})^{n-2}
\nonumber\\
&& \times \underline{x}^\mu \underline{x}^\rho \underline{x}^\lambda \underline{x}^\tau z^\nu z^\sigma G_{\mu\nu;\lambda} G_{\rho\sigma;\tau}, \nonumber\\
(z\cdot \tensor{D})^n_{6(4)} &=& -\frac{n(n-1)}{8} (z\cdot \tensor{\partial})^{n-2}
\nonumber\\
&& \times \underline{x}^\mu \underline{x}^\rho \underline{x}^\lambda \underline{x}^\tau z^\nu z^\sigma G_{\mu\nu;\lambda\tau} G_{\rho\sigma}, \nonumber\\
(z\cdot \tensor{D})^n_{6(5)} &=& \left[ -\frac{i}{72} \times n (z\cdot \tensor{\partial})^{n-1} \underline{x}^\mu \underline{x}^\rho \underline{x}^\sigma \underline{x}^\lambda \underline{x}^\tau z^\nu \right. \nonumber\\
&& - \frac{i}{216} n(n-1)(n-2) (z\cdot \tensor{\partial})^{n-3}
\nonumber\\
&& \times \underline{x}^\mu \left( z^\rho z^\sigma \underline{x}^\lambda \underline{x}^\tau + z^\rho \underline{x}^\sigma z^\lambda \underline{x}^\tau + z^\rho \underline{x}^\sigma \underline{x}^\lambda z^\tau \right. \nonumber\\
&& \left. + \underline{x}^\rho z^\sigma z^\lambda \underline{x}^\tau + \underline{x}^\rho z^\sigma \underline{x}^\lambda z^\tau + \underline{x}^\rho \underline{x}^\sigma z^\lambda z^\tau \right) z^\nu \nonumber\\
&& - \frac{i}{360} n(n-1)(n-2)(n-3)(n-4) \nonumber\\
&& \left. \times (z\cdot \tensor{\partial})^{n-5} \underline{x}^\mu z^\rho z^\sigma z^\lambda z^\tau z^\nu \right] G_{\mu\nu;\rho\sigma\lambda\tau}, \nonumber
\end{eqnarray}
where the subscript $k(m)$ with $k=(1,\cdots,6)$ stands for the dimension of the operator, in which $(m)$ stands for the $m$-th type of the operator with same dimension. For example, there is two type of dimension four operators, three type of dimension five operators and five type of dimension-six operators. Fig.(\ref{vertex}) represents the Feynman diagrams for the vertex operator $\Gamma (z\cdot \tensor{D})^n$ under the background field theory, where thirteen figures, i.e. Fig.(\ref{vertex}a), $\cdots$, Fig.(\ref{vertex}m), correspond to $\Gamma(z\cdot \tensor{D})^n_1$, $\cdots$, $\Gamma(z\cdot \tensor{D})^n_{6(5)}$, respectively. The cross ($\times$) attached to the gluon line indicates the tensor of the local gluon background field with ``$n$" stands for the $n$-th order covariant derivative.

\section{pion leading-twist DA within the QCD sum rules}

\begin{figure}[htb]
\includegraphics[width=0.45\textwidth]{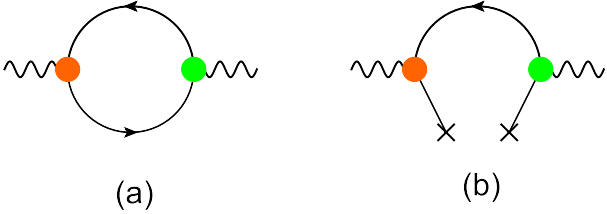}
\caption{Schematic Feynman diagrams for the pion leading-twist DA moments, where the cross $(\times)$ stands for the background quark field. The left big dot and the right big dot stand for the vertex operators ${z\!\!\!\slash}\gamma_5 (z\cdot \tensor{D})^n$ and ${z\!\!\!\slash}\gamma_5$ in the correlator, respectively. The full Feynman diagrams can be found in Figs.(\ref{feyna},\ref{feynb}). }
\label{feyn}
\end{figure}

\begin{figure*}
\includegraphics[width=0.85\textwidth]{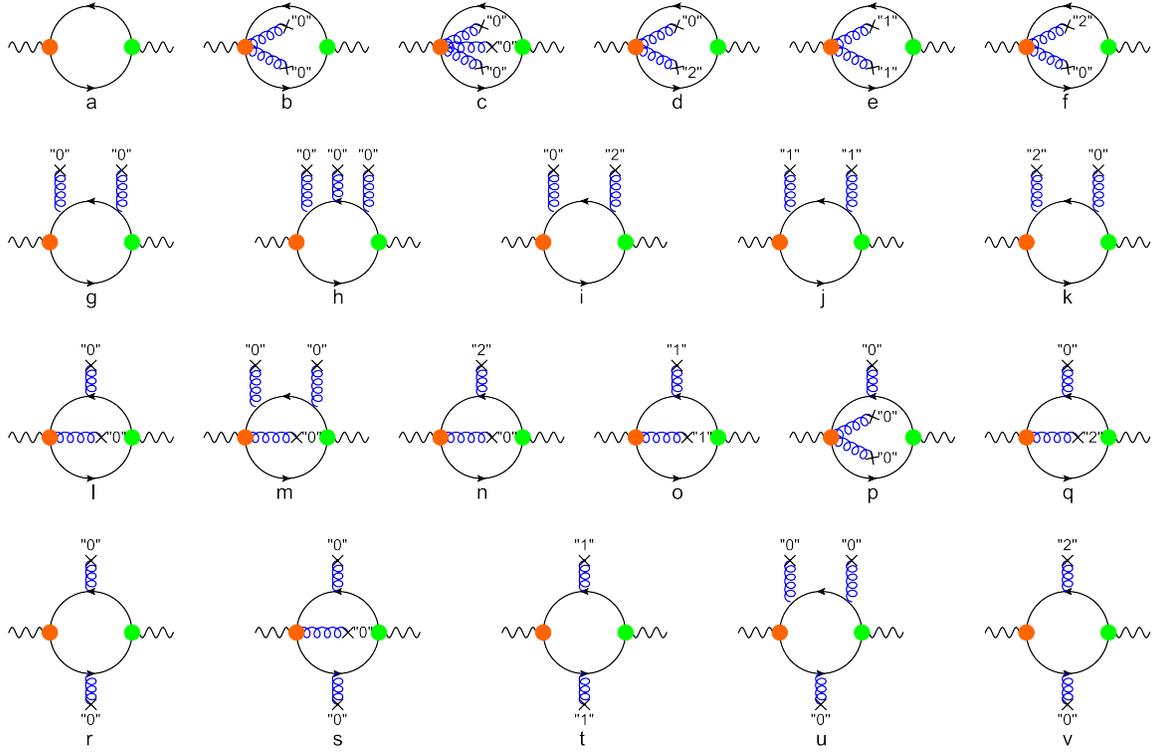}
\caption{The sub-diagrams of Fig.(\ref{feyn}a), where those diagrams that can be obtained by permutation from the asymmetrical diagrams have been omitted. }
\label{feyna}
\end{figure*}

\begin{figure*}
\includegraphics[width=0.85\textwidth]{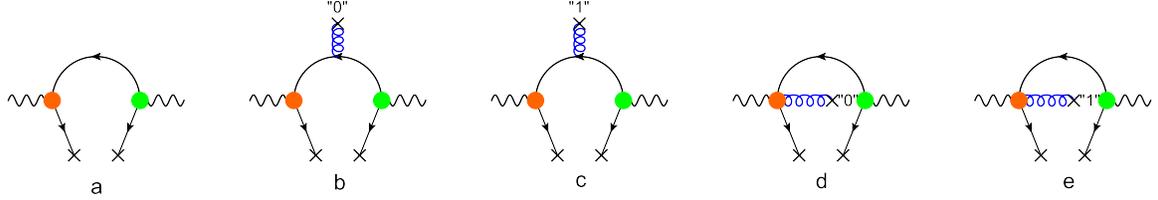}
\caption{The sub-diagrams of Fig.(\ref{feyn}b). }
\label{feynb}
\end{figure*}

Considering the definition
\begin{eqnarray}
\left<0\left| \bar{d}(0) {z\!\!\!\slash} \gamma_5 (iz\cdot \tensor{D})^n u(0) \right|\pi(q)\right> = i(z\cdot q)^{n+1} f_\pi \left<\xi^n\right>,
\label{mom}
\end{eqnarray}
where $f_\pi$ is the pion decay constant and the $n$-th order moments of the pion DA are defined as
\begin{eqnarray}
\left<\xi^n\right> = \int^1_0 du (2u-1)^n \phi_\pi(u). \label{nmom}
\end{eqnarray}
Especially, the $0$-th moment satisfies the normalization condition
\begin{eqnarray}
\left<\xi^0\right> = \int^1_0 du \phi_\pi(u) = 1 . \label{normalize0}
\end{eqnarray}

To derive the sum rules for the pion leading-twist DA moments $\left<\xi^n\right>$, we introduce the following correlation function (correlator),
\begin{eqnarray}
\Pi^{(n,0)}_\pi (z,q) && = i \int d^4x e^{iq\cdot x} \left<0\left| T \left\{ J_n(x) J^\dag_0(0) \right\} \right|0\right>
\nonumber\\ &&
= (z\cdot q)^{n+2} I^{(n,0)}_{\pi} (q^2) ,
\label{correlator}
\end{eqnarray}
where $n=(0,2,4,\cdots)$, $z^2 = 0$ and the currents
\begin{eqnarray}
J_n(x) &=& \bar{d}(x) {z\!\!\!\slash} \gamma_5 (i z\cdot \tensor{D})^n u(x), \\
J^\dagger_0(0) &=& \bar{u}(0) {z\!\!\!\slash} \gamma_5 d(0) .
\end{eqnarray}

In physical region, the correlator (\ref{correlator}) can be treated by inserting a complete set of intermediate hadronic states, which can be simplified with the help of Eq.(\ref{mom}) as
\begin{eqnarray}
\textrm{Im} I^{(n,0)}_{\pi,{\rm had}}(q^2) & =& \pi \delta (q^2 - m_\pi^2) f_\pi^2 \left<\xi^n\right> \nonumber\\
&&+ \pi \frac{3}{4\pi^2 (n+1) (n+3)} \theta (q^2 - s_\pi), \label{hadim}
\end{eqnarray}
where the quark-hadron duality has been adopt and $s_\pi$ stands for the continuum threshold.

On the other hand, we apply the OPE for the correlator (\ref{correlator}) in deep Euclidean region. Substituting the replacement rules (\ref{bfrep0},\ref{bfrep}) into the correlator (\ref{correlator}) and applying the corresponding Feynman rules, we obtain
\begin{eqnarray} &&
\Pi^{(n,0)}_{\pi}(z,q)
= i \int d^4x e^{iq\cdot x}
\nonumber\\ &&
\times \left\{ - \textrm{Tr} \left<0\left| S^d_F(0,x) \not\! z \gamma_5 (iz\cdot \tensor{D})^n S^u_F(x,0) \not\! z \gamma_5 \right|0\right>
\right.
\nonumber\\ &&
+ \textrm{Tr} \left<0\left| \bar{d}(x) d(0) \not\! z \gamma_5 (iz\cdot \tensor{D})^n S^u_F(x,0) \not\! z \gamma_5 \right|0\right>
\nonumber\\ &&
\left.
+ \textrm{Tr} \left<0\left| S^d_F(0,x) \not\! z \gamma_5 (iz\cdot \tensor{D})^n \bar{u}(0) u(x) \not\! z \gamma_5 \right|0\right> \right\}
\nonumber\\ &&
+ \cdots.
\label{ope}
\end{eqnarray}

The Feynman diagrams for the correlator (\ref{ope}) are showed in Fig.(\ref{feyn}), whose details are further presented in Figs.(\ref{feyna},\ref{feynb}). The solid and helical lines are for the quark and gluon propagators, respectively. The cross $(\times)$ symbol stands for the background quark field. The left big dot and the right big dot are the vertex operators ${z\!\!\!\slash}\gamma_5 (z\cdot \tensor{D})^n$ and ${z\!\!\!\slash} \gamma_5$, respectively. Fig.(\ref{feyn}a) corresponds to the first term in Eq.(\ref{ope}), Fig.(\ref{feyn}b) and its permutation-diagrams correspond to the second and third terms of Eq.(\ref{ope}), respectively.

As for Fig.(\ref{feyna}), we observe
\begin{itemize}
\item Figs.(\ref{feyna}l,\ref{feyna}p,\ref{feyna}q,\ref{feyna}s) contribute zero because their quark loop has ${\rm Tr}[\cdots]=0$. The contribution from Fig.(\ref{feyna}g) is negligible due to strong suppression from small $u/d$-quark current masses.
\item Fig.(\ref{feyna}a) provides the perturbative contribution, Figs.(\ref{feyna}b,\ref{feyna}r) provide the contribution proportional to the dimension-four condensate $\left<G^2\right>$, Figs.(\ref{feyna}c,\ref{feyna}h,\ref{feyna}m,\ref{feyna}u) provide the contribution proportional to the dimension-six condensate $\left<g_s^3 f G^3\right>$ and Figs.(\ref{feyna}e,\ref{feyna}j,\ref{feyna}o,\ref{feyna}t) provide the contribution proportional to the dimension-six condensate $g_s^2\sum_{u,d,s}\left<g_s\bar{\psi}\psi\right>^2$. Here $\left<G^2\right>$ and $\left<g_s^3 f G^3\right>$ are abbreviations for $\left<G^A_{\mu\nu}G^{A\mu\nu}\right>$ and $\left<g_s^3 f^{ABC} G^{A\mu\nu}G^{B}_{\ \nu\rho}G^{C\rho}_{\ \ \mu}\right>$.
\item The remaining diagrams involves the contributions proportional to the dimension-six condensate either $\left<g_s^3 f G^3\right>$ or $g_s^2\sum_{u,d,s} \left<g_s\bar{\psi}\psi\right>^2$, where $\sum_{u,d,s}=\sum_{\psi=u,d,s}$.
\end{itemize}

As for Fig.(\ref{feynb}), we observe that
\begin{itemize}
\item Fig.(\ref{feynb}a) provides the contribution proportional to the dimension three condensate $\left<\bar{q}q\right>$ and the dimension five condensate $\left<g_s\bar{q}\sigma TGq\right>$. Fig.(\ref{feynb}b) provides the contribution proportional to either the dimension five condensate $\left<g_s\bar{q}\sigma TGq\right>$ or the dimension-six condensate $\left<g_s\bar{q}q\right>^2$.
\item The remaining three diagrams (\ref{feynb}c,\ref{feynb}d,\ref{feynb}e) provide the contribution proportional to the dimension-six condensate $\left<g_s\bar{q}q\right>^2$, where $q=u,d$.
\end{itemize}

We adopt the $\overline{\rm MS}$-scheme to deal with the infrared divergences emerged in Fig.(\ref{feyna}), whose divergent terms shall be absorbed into the poin DA following the way suggested by Ref.\cite{IRDIV}. During the calculation,we shall meet with the following matrix elements:
\begin{eqnarray}
&& \left<0\left| \bar{q}^a_\alpha(x) q^b_\beta(y) \right|0\right>,\; \left<0\left| \bar{q}^a_\alpha(x) q^b_\beta(y) G^A_{\mu\nu} \right|0\right>,\nonumber\\
&& \left<0\left| \bar{q}^a_\alpha(x) q^b_\beta(y) G^A_{\mu\nu;\rho} \right|0\right>,\; \left<0\left| G^A_{\mu\nu}G^B_{\rho\sigma} \right|0\right>,\nonumber\\
&& \left<0\left| G^A_{\mu\nu}G^B_{\rho\sigma} G^C_{\lambda\tau} \right|0\right>,\; \left<0\left| G^A_{\mu\nu;\lambda}G^B_{\rho\sigma;\tau} \right|0\right>, \nonumber\\
&& \left<0\left| G^A_{\mu\nu}G^B_{\rho\sigma;\lambda\tau} \right|0\right>,\; \left<0\left| G^A_{\mu\nu;\lambda\tau}G^B_{\rho\sigma} \right|0\right>.
\end{eqnarray}
The formulas relating the first three vacuum matrix elements with the conventional condensates can be found in Ref.\cite{ZHONG2}, and the corresponding formulas for the remaining vacuum matrix elements under the $D$-dimensional space, $D = 4-2\epsilon$, are presented in Appendix B.

As a combination the correlator within different regions, the sum rules for the pion DA moments can be derived by the dispersion relation
\begin{eqnarray}
\frac{1}{\pi} \frac{1}{M^2} \int ds e^{-s/M^2} \textrm{Im} I_{\rm had}(s) = \hat{L}_M I_{\rm qcd}(q^2), \label{bordisrel}
\end{eqnarray}
where $M$ is the Borel parameter and the Borel transformation operator
\begin{eqnarray}
\hat{L}_M = \lim_{\begin{array}{c} -q^2,n\to\infty\\ -q^2/n=M^2 \end{array}} \frac{1}{(n-1)!} (-q^2)^n \left( -\frac{d}{d(-q^2)} \right)^n .
\end{eqnarray}

Our final result reads
\begin{widetext}
\begin{eqnarray}
\left<\xi^n\right> && = \frac{M^2 e^{m_\pi^2/M^2}}{f_\pi^2} \left\{ \frac{3}{4\pi^2(n+1)(n+3)} \left( 1 - e^{-s_\pi/M^2} \right) + \frac{m_d \left<\bar{d}d\right> + m_u \left<\bar{u}u\right>}{M^4} + \frac{1}{12\pi} \frac{\left<\alpha_s G^2\right>}{M^4}
\right.
\nonumber\\ &&
- \frac{8n+1}{18} \frac{m_d \left<g_s \bar{d}\sigma TGd\right> + m_u \left<g_s \bar{u}\sigma TGu\right>}{M^6} + \frac{\mathbb{A}}{(4\pi)^2} \frac{\left<g_s^3fG^3\right>}{M^6} + \frac{\mathbb{B}}{(4\pi)^2} \frac{g_s^2 \sum_{u,d,s} \left<g_s\bar{\psi}\psi\right>^2}{M^6}
\nonumber\\ &&
\left.
+ \frac{2(2n+1)}{81} \frac{\left<g_s\bar{d}d\right>^2 + \left<g_s\bar{u}u\right>^2}{M^6} \right\},
\label{momsr}
\end{eqnarray}
where
\begin{eqnarray} &&
\mathbb{A} = \frac{6n+7}{72} - \frac{n}{24} \ln \left(\frac{M^2}{\mu^2}\right) + \theta(n-2) \left[ \frac{7n-3}{72} - \frac{n}{8} \ln \left(\frac{M^2}{\mu^2}\right) + \sum^{n-2}_{n=0} (-1)^k \frac{-2k^2 + 2nk - 2k + 3n}{24 (n-k) (n-k-1)} \right],
\nonumber\\ &&
\mathbb{B} = \frac{8(165n^2 + 323n + 131)}{729(n+1)} + \frac{8(51n+44)}{243} \ln \left(\frac{M^2}{\mu^2}\right) + \theta(n-2) \left\{ -\frac{4(16n^3-89n^2-39n-30)}{729n(n+1)} + \frac{392n}{243} \ln \left(\frac{M^2}{\mu^2}\right)
\right.
\nonumber\\ &&
\quad\quad \left.
+ \sum^{n-2}_{k=0} (-1)^k \left[ \frac{64n(k+1)}{243(n+1)(n-k)} - \frac{8(10k^2-10nk+58k-7n+48)}{243(n-k)(n-k-1)} + \frac{8(8k-5n-2)}{243(k+1)(n-k+1)} \right] \right\}.
\nonumber
\end{eqnarray}
\end{widetext}

As a final remark, we can obtain the Gegenbauer moments of the pion leading-twist DA with the help of $\left<\xi^n\right>$. That is, by substituting Eq.(\ref{DAgegexp}) into Eq.(\ref{nmom}), we have
\begin{eqnarray} &&
a^\pi_2 = \frac{7}{12} \left( 5\left<\xi^2\right> - 1 \right),  \label{gegmom1}\\
&& a^\pi_4 = -\frac{11}{24} \left( 14\left<\xi^2\right> - 21\left<\xi^4\right> - 1 \right), \label{gegmom2}\\
&& a^\pi_6 = \frac{5}{64} \left( 135\left<\xi^2\right> - 495\left<\xi^4\right> + 429\left<\xi^6\right> - 5 \right),  \label{gegmom3} \\ &&
\cdots \cdots \nonumber
\end{eqnarray}

\section{Numerical analysis}

To do the numerical calculation, we adopt~\cite{PDG}: $m_\pi = 139.57018 \pm 0.00035 \textrm{MeV}$ and $f_\pi = 130.41 \pm 0.20 \textrm{MeV}$. As for the condensates,
\begin{itemize}
\item We adopt
\begin{eqnarray}
&& m_u \left<\bar{u}u\right> + m_d \left<\bar{d}d\right> \nonumber\\
&& \simeq - \frac{ f_\pi^2 m_\pi^2 }{2} = (1.656 \pm 0.005) \times 10^{-4} \textrm{GeV}^4.
\label{mqq}
\end{eqnarray}
which is derived from the current algebra~\cite{CURALG} and the partially conserved axial current~\cite{PCAC}.

\item  We adopt~\cite{QUACON},
\begin{eqnarray} &&
m_u \left<g_s\bar{u}\sigma TGu\right> + m_d \left<g_s\bar{d}\sigma TGd\right>
\nonumber\\ && \quad\quad\quad
= (1.325 \pm 0.033) \times 10^{-4} \textrm{GeV}^4 , \label{mqgq}
\end{eqnarray}
which is derived by using $\left<g_s\bar{q}\sigma TGq\right> = m_0^2 \left<\bar{q}q\right>$ with $m_0^2 = 0.80 \pm 0.02 \textrm{GeV}^2$~\cite{m0ref}.

\item As suggested in Ref.\cite{RATQUAGLU}, we adopt~\cite{RQG1,RQG2}
\begin{eqnarray}
\frac{\left<\alpha_s G^2\right>}{\rho \ \alpha_s \left<\bar{u}u\right>^2} = (106 \pm 12) \textrm{GeV}^{-2}, \label{ratquaglu}
\end{eqnarray}
where $\rho \simeq 2 - 3$~\cite{RQG1,RHO1,RHO2,RHO3}. More definitely, we take~\cite{SRREV}
\begin{equation}
\left<\alpha_s G^2\right> = 0.038 \pm 0.011 \textrm{GeV}^4, \label{gg}
\end{equation}
and $\left<\bar{q}q\right> = (-0.24 \pm 0.01)^3 \textrm{GeV}^3$. And we take $\left<g_s\bar{q}q\right>^2 = (1.8 \pm 0.7) \times 10^{-3} \textrm{GeV}^6$.

\item Combining (\ref{relgluqua}) with (\ref{ratquaglu}), together with the ratio $\left<\bar{s}s\right>/ \left<\bar{q}q\right> = 0.74 \pm 0.03$~\cite{QUACON}, we obtain
\begin{eqnarray}
\left<g_s^3fG^3\right> = 0.013 \pm 0.007 \textrm{GeV}^6.
\label{ggg}
\end{eqnarray}
and
\begin{eqnarray}
g_s^2 \sum_{u,d,s} \left<g_s\bar{\psi}\psi\right> = 0.044 \pm 0.024 \textrm{GeV}^6.
\label{gq44}
\end{eqnarray}
\end{itemize}

The leading-order $\alpha_s$ is fixed by $\alpha_s(M_Z)=0.1184 \pm 0.0007$~\cite{ALPHAS} and the renormalization scale is taken as $\mu = M$. Usually, the continuum threshold $s_\pi$ is taken as the square of the first exciting state of pion, i.e., $\pi(1300)$, which however may underestimate some continuum states' contributions. At the present, we determine the value of $s_\pi$ from the sum rules of the $0$-th moment $\left<\xi^0\right>$ together with its normalization condition (\ref{normalize0}), which leads to $s_\pi\simeq 1.1\textrm{GeV}^2$.

\begin{table}[htb]
\caption{The moments $\left<\xi^n\right>$ of the pion leading-twist DA at the scale $\mu=M$ and various condensates' contributions over the total dispersion integration. $\left<\bar{q}Gq\right>$ and $\sum \left<g_s^2\bar{\psi}\psi\right>^2$ are abbreviations for $\left<g_s \bar{q}\sigma TGq\right>$ and $g_s^2 \sum_{u,d,s} \left<g_s\bar{\psi}\psi\right>^2$.}
\begin{center}
\begin{tabular}{c c c c}
\hline
~$n$~ & ~$2$~ & ~$4$~ & ~$6$~ \\
\hline
~$M^2(\textrm{GeV}^2)$~ & ~$[1.231,1.490]$~ & ~$[1.697,1.824]$~ & ~$[2.066,2.187]$~ \\
~$\left<\xi^n\right>$~ & ~$0.331 \pm 0.009$~ & ~$0.198 \pm 0.005$~ & ~$0.146 \pm 0.004$~\\
~$m_q\left<\bar{q}q\right>$~ & ~$-(1.2-1.6)\%$~ & ~$-(1.7-1.8)\%$~ & ~$-(1.9-2.0)\%$~ \\
~$\left<\alpha_s G^2\right>$~ & ~$(7.5-9.9)\%$~ & ~$(10-11)\%$~ & ~$(12-12)\%$~ \\
~$m_q\left<\bar{q}Gq\right>$~ & ~$(0.6-1.0)\%$~ & ~$(1.4-1.6)\%$~ & ~$(1.9-2.2)\%$~ \\
~$\left<g_s^3fG^3\right>$~ & ~$(0.2-0.4)\%$~ & ~$(0.5-0.6)\%$~ & ~$\sim 0.7\%$~ \\
~$\sum \left<g_s^2\bar{\psi}\psi\right>^2$~ & ~$(6.9-11)\%$~ & ~$(12-14)\%$~ & ~$(15-17)\%$~ \\
~$\left<g_s\bar{q}q\right>^2$~ & ~$(2.2-3.6)\%$~ & ~$(4.4-5.2)\%$~ & ~$(6.1-6.8)\%$~ \\
\hline
\end{tabular}
\label{tabxin}
\end{center}
\end{table}

We adopt the usual criteria to fix the Borel window for the moments $\left<\xi^n\right>$, i.e. the continuum state's contribution is less than $40\%$ of the total dispersion integration and the dimension-six condensate's contribution does not exceed $15\%$, $20\%$ and $25\%$ for the second, forth and sixth moments, respectively. We present the moments $\left<\xi^n\right>$ together with their Borel windows in Table \ref{tabxin}. From Table \ref{tabxin}, we observe that the dimension-six condensates, such as $g_s^2 \sum_{u,d,s} \left<g_s\bar{\psi}\psi\right>^2$, do provide sizable contributions to  $\left<\xi^n\right>$, i.e. they are comparable to the lower dimensional condensates' contributions.

\begin{figure}[htb]
\centering
\includegraphics[width=0.5\textwidth]{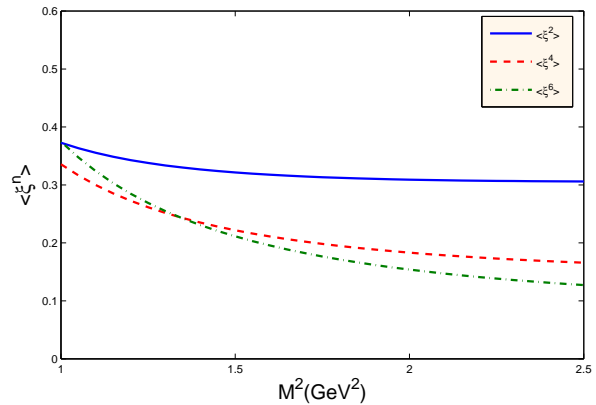}
\caption{The first three moments for the pion leading-twist DA versus the Borel parameter $M^2$.} \label{xin}
\end{figure}

Fig.(\ref{xin}) shows that the first three moments of the pion leading-twist DA versus the Borel parameter $M^2$, where the solid, the dashed and the doted lines are for second, forth and sixth moments, respectively. By taking all uncertainty sources into consideration, and adding the uncertainties in quadrature, we obtain
\begin{eqnarray}
\left< \xi^2 \right>|_{\mu = 1{\rm GeV}} &=& 0.338 \pm 0.032, \label{mom21}\\
\left< \xi^4 \right>|_{\mu = 1{\rm GeV}} &=& 0.211 \pm 0.030, \label{mom22} \\
\left< \xi^6 \right>|_{\mu = 1{\rm GeV}} &=& 0.163 \pm 0.030, \label{mom23}
\end{eqnarray}
whose errors are dominated by the uncertainties of the condensates $\left<\alpha_s G^2\right>$, $\left<g_s\bar{q}q\right>^2$ and $g_s^2 \sum_{u,d,s} \left<g_s\bar{\psi}\psi\right>$. The present results for $\mu=1\;{\rm GeV}$ are obtained by first get the DA behavior with the help of the moments at $\mu=M$ and apply the pion DA running equation~\cite{ASDA} to run the DA from $\mu=M$ to $\mu=1$ GeV, and then get the moments at $\mu=1\;{\rm GeV}$.

\begin{table}[htb]
\caption{The second Gegenbauer moment $a^\pi_2(1\;{\rm GeV})$ under the QCD sum rules and the lattice QCD, respectively. When $a^\pi_2$ is given with different scale other than $1$ GeV, it shall be evolved to $1$ GeV by using the pion DA leading-order running behavior~\cite{ASDA}.}
\begin{center}
\begin{tabular}{c| c c c}
\hline
~~ & ~$a^\pi_2(1\textrm{GeV})$~ & ~Reference~ \\
\hline
~QCD SR~ & ~$0.403 \pm 0.093$~ & ~\textrm{this paper}~ \\
~~ & ~$0.56$~ & ~\cite{CZDA}~ \\
~~ & ~$0.26^{+0.21}_{-0.09}$~ & ~\cite{PDA_SR4}~ \\
~~ & ~$0.28 \pm 0.08$~ & ~\cite{PDA_SR3}~ \\
~~ & ~$0.19 \pm 0.06$~ & ~\cite{PDA_SR5}~ \\
\hline
~LQCD~ & ~$0.381 \pm 0.234^{+0.114}_{-0.062}$~ & ~\cite{PDA_L3}~ \\
~~ & ~$0.364 \pm 0.126$~ & ~\cite{PDA_L4}~ \\
\hline
\end{tabular}
\label{taba2}
\end{center}
\end{table}

\begin{table*}[htb]
\caption{Typical pion DA parameters at $\mu =1{\rm GeV}$.}
\begin{tabular}{ c | c c c | c c c | c c c}
\hline\hline
~$a^\pi_2$~& ~$$~ & ~$0.310$~ & ~$$~& ~$$~ & ~$0.403$~ & ~$$~& ~$$~ & ~$0.496$~ & ~$$~ \\
~$a^\pi_4$~& ~$0.237$~ & ~$0.320$~ & ~$0.403$~& ~$0.237$~ & ~$0.320$~ & ~$0.403$~& ~$0.237$~ & ~$0.320$~ & ~$0.403$~ \\
\hline
~$A_\pi$~& ~$19.297$~ & ~$18.874$~ & ~$18.477$~& ~$18.291$~ & ~$17.912$~ & ~$17.545$~& ~$17.348$~ & ~$17.003$~ & ~$16.666$~ \\
~$B_2$~& ~$0.211$~ & ~$0.195$~ & ~$0.176$~& ~$0.290$~ & ~$0.271$~ & ~$0.251$~& ~$0.371$~ & ~$0.351$~ & ~$0.329$~ \\
~$B_4$~& ~$0.248$~ & ~$0.322$~ & ~$0.394$~& ~$0.239$~ & ~$0.313$~ & ~$0.385$~& ~$0.229$~ & ~$0.303$~ & ~$0.376$~ \\
~$\beta_\pi(\textrm{GeV})$~& ~$0.708$~ & ~$0.725$~ & ~$0.741$~& ~$0.730$~ & ~$0.746$~ & ~$0.762$~& ~$0.750$~ & ~$0.766$~ & ~$0.783$~ \\
\hline\hline
\end{tabular}
\label{DAparameter}
\end{table*}

Using those moments (\ref{mom21},\ref{mom22},\ref{mom23}), together with the formulas (\ref{gegmom1},\ref{gegmom2},\ref{gegmom3}), we can obtain the Gegenbauer moments $a^{\pi}_{2}$, $a^{\pi}_{4}$ and $a^{\pi}_{6}$. For example, we obtain $a_2^\pi|_{\rm 1\;GeV} = 0.403 \pm 0.093$ and $a_4^\pi|_{\rm 1\;GeV}=0.320\pm0.083$. We present a comparison of $a_2^\pi(1\textrm{GeV})$ under the SVZ sum rules and the lattice QCD in Tab.\ref{taba2}. Phenomenologically, a LCSR analysis on the pion electromagnetic form factor gives $a_2^\pi(1\;{\rm GeV}) = 0.24 \pm 0.14 \pm 0.08$~\cite{PDA_LC3} and $0.20 \pm 0.03$~\cite{PDA_LC4}. A LCSR analysis on the pion transition form factor gives $a_2^\pi(1\;{\rm GeV}) = 0.24 \pm 0.14 \pm 0.08$~\cite{PDA_LC3}, $0.19 \pm 0.05$~\cite{PDA_LC1}, $0.32$~\cite{PDA_LC21}, $0.44$~\cite{PDA_LC22} and $0.27$~\cite{PDA_LC23}. A LCSR analysis on the $B\to\pi l\nu$ gives $a_2^\pi(1\;{\rm GeV}) =0.267 \pm 0.228$~\cite{INVDA1}, $0.19 \pm 0.19 \pm 0.08$~\cite{PDA_LC5} and $0.17^{+0.15}_{-0.17}$~\cite{PDA_LC6}. It shows that our present estimation of $a^\pi_2$ agrees with most of the estimations within errors derived in the literature. Especially, our central value is close to the Lattice QCD estimations~\cite{PDA_L3,PDA_L4}.

Following the idea of Ref.\cite{INVDA1}, by including the correlation effect from the transverse distribution, we can construct a pion DA model as
\begin{widetext}
\begin{eqnarray}
\phi_\pi(\mu,x) = \frac{\sqrt{3} A_\pi m_q \beta_\pi}{2\pi^{3/2}f_\pi} \sqrt{x(1-x)} \varphi_\pi(x) \times \left\{ \textrm{Erf}\left[ \sqrt{\frac{m_q^2 + \mu^2}{8\beta_\pi^2 x(1-x)}} \right] - \textrm{Erf} \left[ \sqrt{\frac{m_q^2}{8\beta_\pi^2 x(1-x)}} \right] \right\}, \label{DA_model}
\end{eqnarray}
\end{widetext}
where the light constitute quark $m_q\simeq0.30$ GeV, the error function $\textrm{Erf}(x) = \frac{2}{\sqrt{\pi}} \int^x_0 e^{-t^2} dt$, and the longitudinal part can be constructed as
\begin{displaymath}
\varphi_\pi(x) = \left[1 + B_2 \times C^{3/2}_2(2x-1) + B_4 \times C^{3/2}_4(2x-1)\right].
\end{displaymath}
The parameters $A_\pi$, $\beta_\pi$, $B_2$ and $B_4$ can be fixed by the pion DA normalization, the constraint from $\pi\to\gamma\gamma$~\cite{BHL} and the above determined Gegenbauer moments $a^\pi_2$ and $a^\pi_4$. Several typical values for those parameters are presented in Table \ref{DAparameter}.

\section{Summary}

The background field theory provides an unambiguous physical picture for both the QCD physical vacuum and the SVZ sum rules. Under this framework, the OPE of the correlator can be calculated systematically. For the first time, we provide the quark propagator and the vertex up to dimension-six operators under the background field theory. It is noted that those newly dimension-six terms for the quark propagator are important for a sound estimation of a physics process up to dimension-six condensates. Because the mass terms are kept explicitly in the quark propagator, it is applicable for both the massless and massive cases.

In the present paper, we have studied the pion leading-twist DA within the QCD sum rules under the background field theory. The resultant sum rules for the DA moments up to dimension-six condensates have been presented. It has been found that the dimension-six condensates can provide sizable contributions to the DA moments in comparison to contributions from the lower dimension condensates. Within errors, our estimation of $a^\pi_2|_{\rm 1\;GeV} = 0.403 \pm 0.093$ agrees with most of the estimations derived in the literature. Its central value is larger than those determined by previous QCD sum rules, which is mainly caused by the contribution from the dimension-six condensate $g_s^2 \sum_{u,d,s} \left<g_s\bar{\psi}\psi\right>^2$.

\begin{figure}[tb]
\centering
\includegraphics[width=0.50\textwidth]{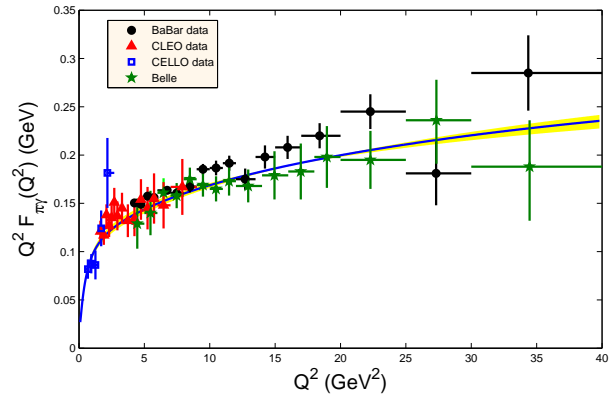}
\caption{$Q^2 F_{\pi\gamma}(Q^2)$ by varying the second and four Gegenbauer moments within their reasonable region, i.e. $a_2^\pi|_{\rm 1\;GeV} = 0.403 \pm 0.093$ and $a_4^\pi|_{\rm 1\;GeV}=0.320\pm0.083$. The solid line stands for the central values for $Q^2 F_{\pi\gamma}(Q^2)$ and the shaded band stands for the theoretical uncertainties caused by the Gegenbauer moments. } \label{fig_PPTFF}
\end{figure}

As an application of the suggested pion DA model (\ref{DA_model}), we redraw the pion-photon TFF $F_{\pi\gamma}(Q^2)$ in Fig.(\ref{fig_PPTFF}). All the formulas for $F_{\pi\gamma}(Q^2)$ can be found in our previous paper \cite{INVDA1}, only one need to change the pion DA used there to be our present model. Fig.(\ref{fig_PPTFF}) indicates that in the large $Q^2$ region, the pion-photon TFF $F_{\pi\gamma}(Q^2)$ lies in between the Belle data and the BABAR data, which is consistent with the recent $k_T$ factorization estimation with the joint resummation~\cite{hnli}.

\vspace{0.5cm}

{\bf Acknowledgments}:
This work was supported in part by the Natural Science Foundation of China under Grant No.11075225, No.11235005 and No.11275280.

\appendix

\section{The full quark propagator in the background field}

Within the framework of the background field theory, the full quark propagator with various gauge invariant tensors that shall result in up to dimension-six operators can be written as
\begin{widetext}
\begin{equation}
S_F(x,0)=S_F^0(x,0)+S_F^2(x,0)+S_F^3(x,0)+ \sum_{i=1}^{2} S_F^{4(i)}(x,0) + \sum_{i=1}^{3} S_F^{5(i)}(x,0) + \sum_{i=1}^{5} S_F^{6(i)}(x,0),
\end{equation}
\end{widetext}
in which the propagators with various gauge invariant tensors are
\begin{widetext}
\begin{eqnarray} &&
S_F^0(x,0) = i \int \frac{d^4p}{(2\pi)^4} e^{-ip\cdot x} \left\{ -\frac{m + \not\! p}{m^2-p^2} \right\} , \label{appA1} \\ &&
S_F^2(x,0) = i \int \frac{d^4p}{(2\pi)^4} e^{-ip\cdot x} \left\{ -\frac{i}{2} \frac{\gamma^\mu (m - \not\! p) \gamma^\nu}{(m^2-p^2)^2} G_{\mu\nu} \right\},\\ &&
S_F^3(x,0) = i \int \frac{d^4p}{(2\pi)^4} e^{-ip\cdot x} \left\{ -\frac{2}{3} \left[ \frac{(\gamma^\mu p^\rho + \gamma^\rho p^\mu) (m - \not\! p)}{(m^2-p^2)^3} - \frac{g^{\mu\rho}}{(m^2-p^2)^2} \right] \gamma^\nu G_{\mu\nu;\rho} \right\},\\ &&
S_F^{4(1)}(x,0) = i \int \frac{d^4p}{(2\pi)^4} e^{-ip\cdot x} \left\{ \left[ \frac{1}{4} \left( \frac{\gamma^\mu (m - \not\! p)}{(m^2-p^2)^3} - 2\frac{p^\mu}{(m^2-p^2)^3} \right) \gamma^\nu \gamma^\rho \gamma^\sigma \right.\right. \nonumber\\ &&
\quad\quad\quad\quad\ \ \left.\left.
+ \frac{1}{2} \left( \frac{(m + \not\! p) \gamma^\mu}{(m^2-p^2)^3} g^{\nu\sigma} + 4\frac{\gamma^\mu (m - \not\! p)}{(m^2-p^2)^4} p^\nu p^\sigma \right) \gamma^\rho \right] G_{\mu\nu}G_{\rho\sigma} \right\} ,  \\ &&
S_F^{4(2)}(x,0) = i \int \frac{d^4p}{(2\pi)^4} e^{-ip\cdot x} \left\{ \frac{i}{4} \left[ \frac{g^{\{\mu\rho}\gamma^{\sigma\}} (m - \not\! p)}{(m^2-p^2)^3} - 2\frac{g^{\{\mu\rho}p^{\sigma\}}}{(m^2-p^2)^3} + 4\frac{\gamma^{\{\mu}p^\rho p^{\sigma\}} (m - \not\! p)}{(m^2-p^2)^4} \right] \gamma^\nu G_{\mu\nu;\rho\sigma} \right\} ,\\ &&
S_F^{5(1)}(x,0) = i \int \frac{d^4p}{(2\pi)^4} e^{-ip\cdot x} \left\{ -\frac{i}{3} \left[ \left( 3\frac{\gamma^\mu (m-\not\! p) \gamma^\nu}{(m^2-p^2)^4} (p^\lambda \gamma^\rho + p^\rho \gamma^\lambda) - \frac{\gamma^\nu (g^{\mu\lambda} \gamma^\rho + g^{\mu\rho} \gamma^\lambda)}{(m^2-p^2)^3} \right) \gamma^\sigma \right.\right.
\nonumber\\ &&
\quad\quad\quad\quad\ \
\left.\left.
+ 4 \left( \frac{\gamma^\mu (m-\not\! p)}{(m^2-p^2)^4} g^{\{\nu\sigma} p^{\lambda\}} + 2 \frac{p^\mu g^{\{\nu\sigma} p^{\lambda\}}}{(m^2-p^2)^4} + 6\frac{\gamma^\mu (m-\not\! p)}{(m^2-p^2)^5} p^\nu p^\sigma p^\lambda \right) \gamma^\rho \right] G_{\mu\nu}G_{\rho\sigma;\lambda} \right\},  \\ &&
S_F^{5(2)}(x,0) = i \int \frac{d^4p}{(2\pi)^4} e^{-ip\cdot x} \left\{ \frac{2i}{3} \left[ \left( \frac{g^{\mu\lambda}}{(m^2-p^2)^3} + \frac{6p^\mu p^\lambda}{(m^2-p^2)^4} \right) \gamma^\nu \gamma^\rho \gamma^\sigma \right.\right. \nonumber\\ &&
\quad\quad\quad\quad\ \ \left.\left.
- 2 \left( \frac{\gamma^\mu (m-\not\! p)}{(m^2-p^2)^4} g^{\{\nu\sigma} p^{\lambda\}} + 2 \frac{p^\mu g^{\{\nu\sigma} p^{\lambda\}}}{(m^2-p^2)^4} + 6\frac{\gamma^\mu (m-\not\! p)}{(m^2-p^2)^5} p^\nu p^\sigma p^\lambda \right) \gamma^\rho \right] G_{\mu\nu;\lambda}G_{\rho\sigma} \right\},  \\ &&
S_F^{5(3)}(x,0) = i \int \frac{d^4p}{(2\pi)^4} e^{-ip\cdot x} \left\{ \frac{4}{15} \left[ \frac{g^{\{\rho\sigma} p^\lambda \gamma^{\mu\}} (m-\not\! p)}{(m^2-p^2)^4} - 2 \frac{g^{\{\rho\sigma} p^\lambda p^{\mu\}}}{(m^2-p^2)^4} + 6 \frac{\gamma^{\{\mu} p^\rho p^\sigma p^{\lambda\}} (m-\not\! p)}{(m^2-p^2)^5} \right.\right.
\nonumber\\ &&
\quad\quad\quad\quad\ \ \left.\left.
- \frac{g^{(\mu\rho\sigma\lambda)}}{(m^2-p^2)^3} \right] \gamma^\nu G_{\mu\nu;\rho\sigma\lambda} \right\},  \\ &&
S_F^{6(1)}(x,0) = i \int \frac{d^4p}{(2\pi)^4} e^{-ip\cdot x} \frac{i}{8} \left\{ \left[ \frac{\gamma^\mu (m - \not\! p)}{(m^2-p^2)^4} - 4\frac{p^\mu}{(m^2-p^2)^4} \right] \gamma^\nu \gamma^\rho \gamma^\sigma \gamma^\lambda \gamma^\tau
\right. \nonumber\\ &&
\quad\quad\quad\quad\ \ \left.
+ 2\left[ 3\frac{\gamma^\mu (m - \not\! p)}{(m^2-p^2)^4} g^{\sigma\tau} + 16\frac{\gamma^\mu (m - \not\! p)}{(m^2-p^2)^5} p^\sigma p^\tau - 4\frac{g^{\mu\sigma}p^\tau + g^{\mu\tau}p^\sigma}{(m^2-p^2)^4} \right] \gamma^\nu \gamma^\rho \gamma^\lambda \right\} G_{\mu\nu}G_{\rho\sigma}G_{\lambda\tau} , \\ &&
S_F^{6(2)}(x,0) = i \int \frac{d^4p}{(2\pi)^4} e^{-ip\cdot x} \left(-\frac{1}{8}\right) \left\{ \left[ 3\frac{\gamma^\mu (m - \not\! p) \gamma^\nu}{(m^2-p^2)^4} g^{\{\lambda\tau} \gamma^{\rho\}} + 16\frac{\gamma^\mu (m - \not\! p) \gamma^\nu}{(m^2-p^2)^5} \gamma^{\{\rho} p^\lambda p^{\tau\}} \right.\right. \nonumber\\ &&
\quad\quad\quad\quad\ \ \left.
- 4\frac{\gamma^\nu}{(m^2-p^2)^4} g^{\mu\{\lambda} p^\tau \gamma^{\rho\}} \right] \gamma^\sigma + 4\left[ \frac{m + \not\! p}{(m^2-p^2)^4} g^{(\nu\sigma\tau\lambda)} + 6\frac{m+\not\! p}{(m^2-p^2)^5} g^{\{\nu\sigma} p^{\tau} p^{\lambda\}}
\right. \nonumber\\ &&
\quad\quad\quad\quad\ \ \left.\left.
+ 48\frac{m + \not\! p}{(m^2-p^2)^6} p^\nu p^\sigma p^\tau p^\lambda \right] \gamma^\mu \gamma^\rho \right\} G_{\mu\nu}G_{\rho\sigma;\lambda\tau} , \\ &&
S_F^{6(3)}(x,0) = i \int \frac{d^4p}{(2\pi)^4} e^{-ip\cdot x} \left( -\frac{2}{9} \right) \left\{ 3\left( \left[ 2\frac{\gamma^\mu (m - \not\! p)}{(m^2-p^2)^5} p^\lambda p^\tau - \frac{g^{\{\mu\lambda} p^{\tau\}}}{(m^2-p^2)^4} - \frac{4p^\mu p^\lambda p^\tau}{(m^2-p^2)^5} \right] \gamma^\nu \gamma^\rho \right.\right.
\nonumber\\ &&
\quad\quad\quad\quad\ \ \left.
+ (\mu \leftrightarrow \lambda) + (\rho \leftrightarrow \tau) + (\mu \leftrightarrow \lambda, \rho \leftrightarrow \tau) \right) \gamma^\sigma + 4\left[ \frac{m + \not\! p}{(m^2-p^2)^4} g^{(\nu\lambda\sigma\tau)} \right. \nonumber\\ &&
\quad\quad\quad\quad\ \ \left.\left.
+ 6\frac{m+\not\! p}{(m^2-p^2)^5} g^{\{\nu\lambda} p^{\sigma} p^{\tau\}} + 48\frac{m + \not\! p}{(m^2-p^2)^6} p^\nu p^\lambda p^\sigma p^\tau \right] \gamma^\mu \gamma^\rho \right\} G_{\mu\nu;\lambda}G_{\rho\sigma;\tau} , \\ &&
S_F^{6(4)}(x,0) = i \int \frac{d^4p}{(2\pi)^4} e^{-ip\cdot x} \left( -\frac{1}{2} \right) \left\{ \left[ \frac{m + \not\! p}{(m^2-p^2)^4} g^{(\nu\tau\lambda\sigma)} + 6\frac{m+\not\! p}{(m^2-p^2)^5} g^{\{\nu\tau} p^{\lambda} p^{\sigma\}}
\right.\right.\nonumber\\
&& \quad\quad\quad\quad\ \ \left.\left.
+ 48\frac{m + \not\! p}{(m^2-p^2)^6} p^\nu p^\tau p^\lambda p^\sigma \right] \gamma^\mu \gamma^\rho - 3\left[ \frac{g^{\{\mu\lambda} p^{\tau\}}}{(m^2-p^2)^4} + \frac{8p^\mu p^\lambda p^\tau}{(m^2-p^2)^5} \right] \gamma^\nu \gamma^\rho \gamma^\sigma \right\} G_{\mu\nu;\lambda\tau}G_{\rho\sigma} ,\\
&& S_F^{6(5)}(x,0) = i \int \frac{d^4p}{(2\pi)^4} e^{-ip\cdot x} \left\{ -\frac{i}{18} \left[ \frac{g^{[\rho\sigma\lambda\tau} \gamma^{\mu]} (m-\not\! p)}{(m^2-p^2)^4} - 4 \frac{g^{[\rho\sigma\lambda\tau} p^{\mu]}}{(m^2-p^2)^4} + 6 \frac{g^{\{\rho\sigma} p^\lambda p^\tau \gamma^{\mu\}} (m-\not\! p)}{(m^2-p^2)^5}
\right.\right. \nonumber\\
&& \quad\quad\quad\quad\ \ \left.\left.
- 12 \frac{g^{\{\rho\sigma} p^\lambda p^\tau p^{\mu\}}}{(m^2-p^2)^5} + 48 \frac{p^{\{\rho} p^\sigma p^\lambda p^\tau \gamma^{\mu\}} (m-\not\!p)}{(m^2-p^2)^6} \right] \gamma^\nu G_{\mu\nu;\rho\sigma\lambda\tau} \right\}, \label{prop2}
\end{eqnarray}
\end{widetext}
where
\begin{widetext}
\begin{eqnarray}
g^{(\mu\nu\rho\sigma)} &=& g^{\mu\nu} g^{\rho\sigma} + g^{\mu\rho} g^{\nu\sigma} + g^{\mu\sigma} g^{\nu\rho}, \nonumber\\
g^{[\mu\nu\rho\sigma} p^{\lambda]} &=& g^{(\mu\nu\rho\sigma)} p^\lambda + g^{(\lambda\nu\rho\sigma)} p^\mu + g^{(\mu\lambda\rho\sigma)} p^\nu + g^{(\mu\nu\lambda\sigma)} p^\rho + g^{(\mu\nu\rho\lambda)} p^\sigma,
\nonumber\\
g^{\{\mu\nu} p^{\rho\}} &=& g^{\mu\nu} p^\rho + g^{\mu\rho} p^\nu + g^{\nu\rho} p^\mu,
\nonumber\\
g^{\{\mu\nu} p^\rho p^{\sigma\}} &=& g^{\mu\nu} p^\rho p^\sigma + g^{\mu\rho} p^\nu p^\sigma + g^{\mu\sigma} p^\nu p^\rho + g^{\nu\rho} p^\mu p^\sigma + g^{\nu\sigma} p^\mu p^\rho + g^{\rho\sigma} p^\mu p^\nu, \nonumber\\
g^{\{\mu\nu} p^\rho p^\sigma p^{\lambda\}} &=& g^{\mu\nu} p^\rho p^\sigma p^\lambda + g^{\mu\rho} p^\nu p^\sigma p^\lambda + g^{\mu\sigma} p^\rho p^\nu p^\lambda + g^{\mu\lambda} p^\rho p^\sigma p^\nu + g^{\nu\rho} p^\mu p^\sigma p^\lambda + g^{\nu\sigma} p^\rho p^\mu p^\lambda \nonumber\\
&& + g^{\nu\lambda} p^\rho p^\sigma p^\mu + g^{\rho\sigma} p^\mu p^\nu p^\lambda + g^{\rho\lambda} p^\mu p^\sigma p^\nu + g^{\sigma\lambda} p^\rho p^\mu p^\nu
\nonumber\\
g^{\{\mu\rho} \gamma^{\sigma\}} &=& g^{\mu\rho} \gamma^\sigma + g^{\mu\sigma} \gamma^\rho + g^{\rho\sigma} \gamma^{\mu}, \nonumber\\
g^{[\rho\sigma\lambda\tau} \gamma^{\mu]} &=& \gamma^\mu g^{(\rho\sigma\lambda\tau)} + (\mu \leftrightarrow \rho) + (\mu \leftrightarrow \sigma) + (\mu \leftrightarrow \lambda) + (\mu \leftrightarrow \tau), \nonumber\\
g^{\{\rho\sigma} p^\lambda \gamma^{\mu\}} &=& \gamma^\mu g^{\{\rho\sigma} p^{\lambda\}} + (\mu \leftrightarrow \rho) + (\mu \leftrightarrow \sigma) + (\mu \leftrightarrow \lambda),
\nonumber\\
g^{\{\rho\sigma} p^\lambda p^\tau \gamma^{\mu\}} &=& \gamma^\mu g^{\{\rho\sigma} p^\lambda p^{\tau\}} + (\mu \leftrightarrow \rho) + (\mu \leftrightarrow \sigma) + (\mu \leftrightarrow \lambda) + (\mu \leftrightarrow \tau), \nonumber\\
\gamma^{\{\mu} p^\rho p^{\sigma\}} &=& \gamma^\mu p^\rho p^\sigma + \gamma^\rho p^\mu p^\sigma + \gamma^\sigma p^\mu p^\rho, \nonumber\\
\gamma^{\{\mu} p^\rho p^\sigma p^{\lambda\}} &=& \gamma^\mu p^\rho p^\sigma p^\lambda + (\mu \leftrightarrow \rho) + (\mu \leftrightarrow \sigma) + (\mu \leftrightarrow \lambda),
\nonumber\\
\gamma^{\{\mu} p^\rho p^\sigma p^\lambda p^{\tau\}} &=& \gamma^\mu p^\rho p^\sigma p^\lambda p^\tau + (\mu \leftrightarrow \rho) + (\mu \leftrightarrow \sigma) + (\mu \leftrightarrow \lambda) + (\mu \leftrightarrow \tau), \nonumber\\
g^{\mu\{\lambda} p^\tau \gamma^{\rho\}} &=& g^{\mu\lambda} p^\tau \gamma^\rho + g^{\mu\tau} p^\lambda \gamma^\rho + g^{\mu\rho} p^\tau \gamma^\lambda + g^{\mu\tau} p^\rho \gamma^\lambda + g^{\mu\rho} p^\lambda \gamma^\tau + g^{\mu\lambda} p^\rho \gamma^\tau.
\nonumber
\end{eqnarray}
\end{widetext}
In those expressions, the mass terms are kept explicitly, which are appropriate not only for the light-quark case but also for the heavy-quark one.

\section{Vacuum matrix elements in the $D$-dimension space}

The vacuum matrix elements in the $D$-dimension space are
\begin{eqnarray} &&
\left<0\left| G^A_{\mu\nu}G^B_{\rho\sigma} \right|0\right>
\nonumber\\ && \quad\quad
= \frac{\left<G^2\right>}{8D(D-1)} \delta^{AB} \left( g_{\mu\rho}g_{\nu\sigma} - g_{\mu\sigma}g_{\nu\rho} \right),
\nonumber\\ &&
\textrm{Tr}_C \left<0\left| G_{\mu\nu}G_{\rho\sigma} \right|0\right>
\nonumber\\ && \quad\quad
= \frac{2\pi}{D(D-1)} \left<\alpha_s G^2\right> \left( g_{\mu\rho}g_{\nu\sigma} - g_{\mu\sigma}g_{\nu\rho} \right),
\nonumber\\ &&
\left<0\left| G^A_{\mu\nu;\lambda} G^B_{\rho\sigma;\tau} \right|0\right>
\nonumber\\ && \quad\quad
= -\frac{2}{27 D^2(D-1)} \sum_{u,d,s} \left<g_s\bar{\psi}\psi\right>^2 \delta^{AB}
\nonumber\\ && \quad\quad
\times \left[ 2 g_{\lambda\tau} (g_{\mu\rho}g_{\nu\sigma} - g_{\mu\sigma}g_{\nu\rho}) + g_{\lambda\rho} (g_{\tau\mu}g_{\sigma\nu} - g_{\tau\nu}g_{\sigma\mu})
\right.
\nonumber\\ && \quad\quad
\left.
- g_{\lambda\sigma} (g_{\tau\mu}g_{\rho\nu} - g_{\tau\nu}g_{\rho\mu}) \right],
\nonumber\\ &&
\textrm{Tr}_C \left<0\left| G_{\mu\nu;\lambda} G_{\rho\sigma;\tau} \right|0\right>
\nonumber\\ && \quad\quad
= -\frac{8}{27D^2(D-1)} g_s^2 \sum_{u,d,s} \left<g_s\bar{\psi}\psi\right>^2
\nonumber\\ && \quad\quad
\times \left[ 2 g_{\lambda\tau} (g_{\mu\rho}g_{\nu\sigma} - g_{\mu\sigma}g_{\nu\rho}) + g_{\lambda\rho} (g_{\tau\mu}g_{\sigma\nu} - g_{\tau\nu}g_{\sigma\mu})
\right.
\nonumber\\ && \quad\quad
\left.
- g_{\lambda\sigma} (g_{\tau\mu}g_{\rho\nu} - g_{\tau\nu}g_{\rho\mu}) \right],
\nonumber\\ &&
\left<0\left| G^A_{\mu\nu} G^B_{\rho\sigma} G^C_{\lambda\tau} \right|0\right>
\nonumber\\ && \quad\quad
= \frac{\left<fG^3\right>}{24 D (D-1) (D-2)} f^{ABC}
\nonumber\\ && \quad\quad
\times \left\{ \left[ g_{\mu\rho} (g_{\nu\lambda}g_{\sigma\tau} - g_{\nu\tau}g_{\sigma\lambda}) - g_{\mu\sigma} (g_{\nu\lambda}g_{\rho\tau} - g_{\nu\tau}g_{\rho\lambda}) \right]
\right.
\nonumber\\ && \quad\quad
\left.
- \left[ g_{\nu\rho} (g_{\mu\lambda}g_{\sigma\tau} - g_{\mu\tau}g_{\sigma\lambda}) - g_{\nu\sigma} (g_{\mu\lambda}g_{\rho\tau} - g_{\mu\tau}g_{\rho\lambda}) \right] \right\},
\nonumber\\ &&
\textrm{Tr}_C \left<0\left| G_{\mu\nu} G_{\rho\sigma} G_{\lambda\tau} \right|0\right>
\nonumber\\ && \quad\quad
= \frac{i}{4 D (D-1) (D-2)} \left<g_s^3fG^3\right>
\nonumber\\ && \quad\quad
\times \left\{ \left[ g_{\mu\rho} (g_{\nu\lambda}g_{\sigma\tau} - g_{\nu\tau}g_{\sigma\lambda}) - g_{\mu\sigma} (g_{\nu\lambda}g_{\rho\tau} - g_{\nu\tau}g_{\rho\lambda}) \right]
\right.
\nonumber\\ && \quad\quad
\left.
- \left[ g_{\nu\rho} (g_{\mu\lambda}g_{\sigma\tau} - g_{\mu\tau}g_{\sigma\lambda}) - g_{\nu\sigma} (g_{\mu\lambda}g_{\rho\tau} - g_{\mu\tau}g_{\rho\lambda}) \right] \right\},
\nonumber\\ &&
\left<0\left| G^A_{\mu\nu} G^B_{\rho\sigma;\lambda\tau} \right|0\right>
\nonumber\\ && \quad\quad
= \delta^{AB} \left\{ \left( a \times \sum_{u,d,s} \left<g_s\bar{\psi}\psi\right>^2 - b \times \left< g_s f G^3 \right> \right)
\right.
\nonumber\\ && \quad\quad
\times \left[ 2 g_{\lambda\tau} (g_{\mu\sigma}g_{\nu\rho} - g_{\mu\rho}g_{\nu\sigma}) + g_{\rho\tau} (g_{\mu\sigma}g_{\nu\lambda} - g_{\mu\lambda}g_{\nu\sigma})
\right.
\nonumber\\ && \quad\quad
\left.
- g_{\sigma\tau} (g_{\mu\lambda}g_{\nu\rho} - g_{\mu\rho}g_{\nu\lambda}) \right]
\nonumber\\ && \quad\quad
+ \left( a \times \sum_{u,d,s} \left<g_s\bar{\psi}\psi\right>^2 + b \times \left< g_s f G^3 \right> \right)
\nonumber\\ && \quad\quad
\left.
\times \left[ g_{\mu\tau} (g_{\rho\nu}g_{\sigma\lambda} - g_{\rho\lambda}g_{\nu\sigma}) + g_{\nu\tau} (g_{\rho\lambda}g_{\sigma\mu} - g_{\rho\mu}g_{\sigma\lambda}) \right] \right\},
\nonumber\\ &&
\textrm{Tr}_C \left<0\left| G_{\mu\nu} G_{\rho\sigma;\lambda\tau} \right|0\right>
\nonumber\\ && \quad\quad
= 4\left( a \times g_s^2 \sum_{u,d,s} \left<g_s\bar{\psi}\psi\right>^2 - b \times \left< g_s^3 f G^3 \right> \right)
\nonumber\\ && \quad\quad
\times \left[ 2 g_{\lambda\tau} (g_{\mu\sigma}g_{\nu\rho} - g_{\mu\rho}g_{\nu\sigma}) + g_{\rho\tau} (g_{\mu\sigma}g_{\nu\lambda} - g_{\mu\lambda}g_{\nu\sigma})
\right.
\nonumber\\ && \quad\quad
\left.
- g_{\sigma\tau} (g_{\mu\lambda}g_{\nu\rho} - g_{\mu\rho}g_{\nu\lambda}) \right]
\nonumber\\ && \quad\quad
+ 4\left( a \times g_s^2 \sum_{u,d,s} \left<g_s\bar{\psi}\psi\right>^2 + b \times \left< g_s^3 f G^3 \right> \right)
\nonumber\\ && \quad\quad
\times \left[ g_{\mu\tau} (g_{\rho\nu}g_{\sigma\lambda} - g_{\rho\lambda}g_{\nu\sigma}) + g_{\nu\tau} (g_{\rho\lambda}g_{\sigma\mu} - g_{\rho\mu}g_{\sigma\lambda}) \right],
\nonumber
\end{eqnarray}
where
\begin{eqnarray}
a = \frac{-2}{9D^2(D-1)(D+2)}, \quad b = \frac{3}{32D^2(D-1)},
\nonumber
\end{eqnarray}
and the symbol $\textrm{Tr}_C$ stands for tracing to the colour matrixes. Obviously,
\begin{eqnarray}
\left<0\left| G^B_{\rho\sigma;\lambda\tau} G^A_{\mu\nu} \right|0\right> = \left<0\left| G^A_{\mu\nu} G^B_{\rho\sigma;\lambda\tau} \right|0\right>.
\nonumber
\end{eqnarray}

Our results agree with those of Refs.\cite{GLUCON1,GLUCON2,GLUCON3}, except for the matrix element $\left<0\left| G^A_{\mu\nu}G^B_{\rho\sigma;\lambda\tau} \right|0\right>$. The result for $\left<0\left| G^A_{\mu\nu}G^B_{\rho\sigma;\lambda\tau} \right|0\right>$ given Ref.\cite{GLUCON3} is incorrect. It is noted that the correct equation for $\widetilde{D}^2 G^A_{\mu\nu}$ is
\begin{eqnarray}
\widetilde{D}^2 G^A_{\mu\nu} &=& \frac{3}{4} g_s f^{ABC} \left[ G^B_{\mu\alpha}G^C_{\nu\alpha} - G^B_{\nu\alpha}G^C_{\mu\alpha} \right]  \nonumber\\
&& + G^A_{\mu\alpha;\alpha\nu}- G^A_{\nu\alpha;\alpha\mu},
\end{eqnarray}
while in Ref.\cite{GLUCON3}, the coefficient of the first term has been taken as $1$.

As a final remark, by making use of the equation
\begin{equation}
\left[ \widetilde{D}_\mu, \widetilde{D}_\nu \right]^{AB} = -g_s f^{ABC} G^C_{\mu\nu},
\end{equation}
we can obtain a useful relation
\begin{equation}
\left<g_s^3 fG^3\right> = \frac{8}{27} g_s^2 \sum_{u,d,s} \left<g_s\bar{\psi}\psi\right>^2. \label{relgluqua}
\end{equation}

\end{document}